\def \beq {\begin{equation}}
\def \bea {\begin{eqnarray}}
\def \eea {\end{eqnarray}}
\def \eeq {\end{equation}}

\newcommand{\nn}{\nonumber}

\documentstyle[11pt,epsfig]{article}

\begin{document}
\begin{titlepage}
%
\begin{center}
%
\hspace*{\fill}{\parbox[t]{1.4in}{
 DESY 98-022  \\
 CERN-TH/98-57  \\
\hfill November 1999
}}
\\[2cm]
{\large\bf A SCREENING MECHANISM}
\\[0.3cm]
{\large\bf FOR EXTRA W AND Z GAUGE BOSONS}
\\[1.3cm]
Joaquim Matias$^{a}$ and Alessandro Vicini$^{b}$
\\[1.4mm]
 {\it a~ CERN Theory Division, 1211 Geneva 23, Switzerland}\\
 {\it b~ DESY, Platanenallee 6, D-15738 Zeuthen, Germany}\\
\end{center}

%
\vspace*{.7cm}

\begin{abstract}

\bigskip

We generalize a previous construction of a fermiophobic model to the
case of more than one extra $W$ and $Z$ gauge bosons. We focus in
particular on the existence of screening configurations and their
implication on the gauge boson mass spectrum. 
One of these configurations allows for the existence of a set of
relatively light new gauge bosons, without violation of the quite restrictive
bounds coming from the $\rho_{\rm NC}$ parameter.
The links with Bess and degenerate Bess models are also discussed.
Also the signal given here by this more traditional gauge extension of
the SM
could help to disentangle it from the towers  of Kaluza-Klein states 
over $W$ and $Z$ gauge bosons in extra dimensions.
\end{abstract}
\vspace{4.5cm}

\leftline{}
\leftline{CERN-TH/98-57}
\hspace*{-1cm}{\rule{10cm}{.1mm}} \\
\vspace{.15cm}
E-mail addresses: Joaquim.Matias@cern.ch \\[-5mm]

~~~~~~~~~~~~~~~~~~~~~vicini@ifh.de

\vspace{1.2cm}
\end{titlepage}
\pagebreak

\setcounter{page}{1}

{\Large{ \bf  Introduction}}
\bigskip

\hspace{5mm}

It has been argued several times in the literature that the Standard
Model (SM) should be considered as a low-energy realization of a more
fundamental theory, but
the lack of evidence for the existence of new particles at LEP1 energies,
apart from those predicted by the SM (with the exception of
the SM Higgs boson), imposes severe restrictions on its extensions.
Moreover, the precision in the measurement of the ratio between the
strength of the neutral and charged current couplings,
the so-called $\rho_{\rm NC}$ parameter (NC stands for neutral current) 
\cite{vel,sir},
rules out models with a too large contribution to this
quantity.

At one loop in perturbation theory the contributions to $\rho_{\rm NC}$
grow
at 
most quadratically in the masses of the scalar particles running in the
loop. The 
quadratic terms can rapidly lead to a disagreement of the prediction with the
experimental data \cite{caral}.
Therefore the phenomenon of ``screening",
i.e. the exact cancellation at one loop
of the contributions to $\rho_{\rm NC}$, which are quadratic in the mass
of the
scalar particles, becomes
an important ingredient to be fulfilled by any viable
alternative to the SM. 
In general the suppression of the new contributions could also occur 
numerically, via a fine-tuning of the parameter of the model, but we will
not 
consider this possibility. We want instead to investigate in detail 
the occurrence of a screening phenomenon, i.e. the conditions to obtain an
exact
analytical cancellation of the terms quadratic in the masses of the scalar
particles.
Then we will analyse the implications of
this requirement in the gauge boson mass spectrum.

An extension of the SM, 
based on an $SU(2)_{L}\times SU(2)_{R} \times U(1)$ gauge group, 
has been presented in a previous paper \cite{FP}.
It has been called fermiophobic (FP) because of the absence of coupling
between the
fermions and the gauge particles of the extra $SU(2)_{R}$  group.
The scalar sector of this model is the minimal one sufficient
to give mass to the vector bosons and to introduce a non-trivial interaction
among them: two doublets and one bidoublet.
No triplet is needed owing to the absence of right-handed neutrinos in
the
model.
Although, in general, $SU(2)$ gauge extensions of the SM are
characterized by large deviations in the $\rho_{\rm NC}$ parameter,
two sets of configurations of the parameter space have been found explicitly
\cite{FP}, which  yield a screening of scalar quadratic contributions.
Moreover, the model in \cite{FP}, apart from being  anomaly-free and
without FCNC at tree level, passes succesfully
all possible phenomenological checks allowing for the existence
of relatively light new gauge bosons. A clear signature of this model
would be the discovery of a relatively light extra $Z^\prime$ mainly
coupled to the hypercharge.

Motivated by the good properties shown by the
fermiophobic model proposed in \cite{FP}
in passing all phenomenological tests, we have studied  a more general
extension, based on an $SU(2)_{L}\times SU(2)^{n}\times U(1)_Y$
(with $n \ge 1$) gauge group, focusing in particular on the existence of
screening configurations and their implications on the gauge boson mass
spectrum.

Although the relation between custodial symmetry and
screening is known \cite{vel,ew},
we have preferred to investigate  the general
expression of $\rho_{\rm NC}$ directly. We have not invoked {\it a
priori} any extra 
global symmetry, but we have studied the conditions under which the
screening
phenomenon occurs and found the corresponding configurations in the parameter
space of the model.
We will see, in detail, how the requirements imposed by these
configurations are so strong that they allow us to get the exact mass
spectrum
of the new gauge particles. The most
remarkable consequence of this model is the
possibility of having  sets of relatively {\bf light} gauge particles
 without contradiction with the experimental data  regarding
$\rho_{\rm NC}$.

In a  series of papers, Casalbuoni et al. \cite{CNL}
have constructed a gauge extension of the SM
that includes,
besides the standard $\gamma$, $Z$ and $W^{\pm}$ gauge vector bosons,
also two new triplets of spin-1 particles. They have
called the model degenerate Bess. The interest of
that model is mainly in its decoupling properties \cite{ACPC}.
The latter originate from an extra
global symmetry that the model has when the gauge couplings are turned
off and that is also responsible for the degeneracy of the vector bosons
masses.
The non-linear realization of this model is able to
reproduce the Higgs-less SM in the limit of infinite
mass of the new vector bosons.
A recently proposed linear realization of this model \cite{CL}
has two interesting limits
depending on the choice of the parameters of the scalar potential:
in one case it
coincides with its  non-linear version (degenerate
Bess) and, for a different choice, with the SM with a 
light Higgs field.
Obviously this implies that at least at low energies the model
shares the phenomenological success of the SM.
We have studied the choices in the parameter space necessary to find a contact
between our model, Bess and degenerate Bess.

Similar extensions of the SM have already been
proposed in the literature \cite{barger}. Some of these extensions
(with $n=1$) appeared in the framework of composite
\cite{Yasue} or supersymmetric models
\cite{langac,senjan}.

More recently, in the context of extra dimensions\cite{emili}, the tower
of Kaluza-Klein states over the W and Z gauge bosons would resemble an
infinite tower of $W^{(n)}$ and $Z^{(n)}$ gauge bosons. The
different type of contributions of these states to $\Delta \rho$ and the
$\epsilon$ parameters as compared to the ones given here for a  more
traditional gauge extension of the SM, could help
to disentangle both cases.

The paper is organized in the following way. In the first section we
present the Lagrangian of the model. In  section 2 we
give the gauge boson mass matrices in the interacting basis for the
charged and neutral sectors where the photon is identified.
We then discuss in section 3 the spectrum of scalars: Goldstones and Higgses.
Section 4 is devoted to the computation of $  \rho_{\rm NC}$.
In section 5 we describe the screening configurations and
the corresponding  mass spectrum for the gauge particles and we also compute
the whole set of $\epsilon$ parameters at tree level for the second
configuration.
We discuss in section 6 the link with degenerate Bess and finally we 
comment on other phenomenological constraints and give our conclusions.

We will adhere in this paper to most of the conventions used in
\cite{FP}.

\section{ \bf The model } \bigskip

We consider a Lagrangian based on a gauge group
$G=SU(2)_{L}\times{SU(2)}^{n}\times U(1)_Y$.
The kinetic and self-interaction terms of the vector bosons are given,
for each sub-group of $G$, by the usual field-strength tensor.
We introduce a complex scalar doublet $\Phi_j$ for each $SU(2)$ group,
and a complex bidoublet $\Psi_{jk}$ for every combination of two distinct
$SU(2)$ groups present in $G$ (where $j,k=0,1,\dots,n$ label the
corresponding groups).
We need at least one complex doublet for each $SU(2)$ group, to give
mass to the new gauge vector bosons, and to break completely $G$ down
to $U(1)_{em}$. We introduce also the bidoublets
because they induce a  non-zero mixing among the different  $SU(2)$ groups of
the theory,
also when these groups are characterized by sensibly different energy
scales. As in the case of $n=1$, triplets are not needed.

This choice  seems to be a good compromise between
the idea of studying the screening in a general kind of scalar sector
and the need of introducing the minimal extension
sufficient to get an acceptable  mass
spectrum for all gauge and fermionic
particles, and also to generate an interaction between the different
$SU(2)$
groups.

The Lagrangian of the Yang-Mills sector is:
\beq
\label{gaulagr}
{\cal L_{{\rm YM}}}~=~
-{1 \over 4} \sum_{j=0}^n \sum_{i=1}^3  G_{j~\mu\nu}^{i} G_j^{i\,
\mu\nu}
 - {1\over 4} B_{\mu\nu} B^{\mu\nu},
\eeq
where
\bea
G_{j~\mu\nu}^i~&=&~\partial_{\mu} W^i_{j~\nu}-\partial_{\nu} W^i_{j~\mu}
 +  g_j \varepsilon^{ilm} W^l_{j~\mu} W^m_{j~\nu}
~~~~~~~i,l,m=1,2,3;~~~~j=0,1,2,\dots,n \nonumber \\
B_{\mu\nu}~&=&~\partial_{\mu} B_{\nu}-\partial_{\nu} B_{\mu}
\eea
are the usual field-strength tensors,
$W^{i}_{j~\mu}$ is the gauge field of the $j^{\rm th}$ $SU(2)$ group,
while $B_{\mu}$ corresponds to the $U(1)_Y$ group.

The quantum numbers of the scalar multiplets are indicated by a list
of $n+2$ elements corresponding to the different groups:
\beq
(0,1,2,\dots,n,Y)
\eeq
and are:
\bea
&&( 1,\dots,\overbrace{2}^{j},\dots,1,-{1 \over 2})~~~~~~~~{\rm for~
each~ doublet~}\Phi_j
    \nonumber \\
&&(1,\dots,\overbrace{2}^{j},\dots,\overbrace{2}^{k},\dots,0) {\rm ~for~
each~
  bidoublet~ }\Psi_{jk} \\
&&~~~~~~~~~~~~~~~~~~~~~~~~~~~~~~~~~j<k~~~{\rm and}~~~j,k~=~0,1,2,\dots,n,
\nonumber
\eea
where 1 and the ellipses stand for a singlet behaviour
with respect to the corresponding $SU(2)$ group of the list and the
2 for a doublet. The values $-{1\over 2}$ and $0$ are the hypercharge
of doublets and bidoublets, respectively.

Starting from these assignments, and therefore from the
transformation rules of each field under a gauge transformation,
we write the covariant derivatives of the scalar multiplets.
\bea
D^{\mu}\Phi_j~&\equiv&~\partial^{\mu}\Phi_j -
i g_j {{{\boldmath{\vec\tau}}}\over 2}\cdot {\vec W_j^{\mu}} \Phi_j +
i {{{\tilde g}}\over 2} B^{\mu} \Phi_j \\
{\rm with}~~\Phi_j~&=&~
\left(
\begin{array}{c}
\phi_j^0 \\
\phi_j^-
\end{array}
\right) \nonumber
\eea
is the covariant derivative of the complex doublet $\Phi_j$, where
${\tilde g}$ is the hypercharge coupling constant,
$g_j$ is the coupling constant of the $j^{\rm th}~SU(2)$ group,
and ${\vec\tau}$ are the Pauli matrices.
\bea
D^{\mu}\Psi_{jk}~&\equiv&~\partial^{\mu}\Psi_{jk}-
i g_j {{{\boldmath{\vec\tau}}}\over 2}\cdot {\vec W_j^{\mu}} \Psi_{jk} +
i g_k \Psi_{jk} {{{\boldmath{\vec\tau}}}\over 2}\cdot {\vec W_k^{\mu}}  \\
{\rm with}~~\Psi_{jk}~&=&~
\left(
\begin{array}{cc}
\varphi^{0 \, (jk)}_{1} & \varphi^{+ \, (jk)}_{2}\\
\varphi^{- \, (jk)}_{1} & \varphi^{0 \, (jk)}_{2}
\end{array}
\right) \nonumber
\eea
is the covariant derivative of the complex bidoublet $\Psi_{jk}$,
$g_j$ and $g_k$ being respectively the coupling constants of the
$j^{\rm th}$ and of the $k^{\rm th}$ $SU(2)$ groups. The ratio between a
gauge
coupling $g_{j}$ and $g_{0}$ will be denoted by
\beq
x_{j}={g_{j} \over g_{0}},
\eeq
where also $x_{0}=1$ and $x_{n+1}\equiv y={\tilde g}/g_{0}$ is introduced
to shorten the
notation in the following.

Starting from the definition of these derivatives, we  explicitly write
the kinetic and the mass terms of the Lagrangian of the scalar sector,
leaving the scalar potential generically indicated with $V(\phi)$.
All the fields are given in the interaction basis:
\bea
\label{scalagr}
{\cal L_{{\rm sca}}}~&=&~
\sum_{j=0}^n \left(D_{\mu}\Phi_j\right)^{\dagger}
                 \left(D^{\mu}\Phi_j\right)
~+~\sum_{j,k=0~j<k}^n {\rm Tr}\left\{\left(D_{\mu}\Psi_{jk}\right)^{\dagger}
                 \left(D^{\mu}\Psi_{jk}\right)\right\}
           ~-~V(\phi).
\eea

We must introduce a gauge-fixing term, and we choose it in the usual
way, in order to remove the mixing terms between Goldstone and gauge
vector bosons:
\bea
\label{gaufix}
{\cal L_{{\rm gf}}}~&=&~
-{1\over {2\xi}}\sum_{j=0}^n \sum_{i=1}^3
\left[
 \partial^{\mu} W^i_{j \, \mu} +
i g_j \xi \left[
\left(\Phi_j'^{\dagger} {{\tau^i}\over 2} \langle \Phi_j
\rangle_0 - \langle \Phi_j^{\dagger} \rangle_0 {{\tau^i}\over 2} \Phi_j'
\right) \right. \right. \nn \\
&&\left.
\left.
+\sum_{k=j+1}^{n}
{\rm Tr}
\left(\Psi_{jk}'^{\dagger} {{\tau^i}\over 2} \langle \Psi_{jk}
\rangle_0 - \langle \Psi_{jk}^{\dagger} \rangle_0 {{\tau^i}\over 2}
\Psi_{jk}' \right)
+\sum_{k=0}^{j-1}
{\rm Tr}
\left(\Psi_{kj}' {{\tau^i}\over 2} \langle \Psi_{kj}^{\dagger}
\rangle_0 - \langle \Psi_{kj} \rangle_0 {{\tau^i}\over 2} \Psi_{kj}'^{
\dagger}
\right) \right]
\right]^2~ \nonumber \\
&&~- {1\over {2\xi}} \sum_{j=0}^{n}
\left[
\partial^{\mu} B_{\mu} -i {{ \tilde g} \, \xi \over 2}
\left(  \Phi_j^{\prime\,\dagger} \langle \Phi_{j} \rangle_{0} -
\langle  \Phi_j^{\dagger} \rangle_{0} \Phi_{j}^{ \prime} \right)
\right]^2 ~
\eea
where
\beq
\langle \Phi_j \rangle_0 ~=~
{1\over{\sqrt{2}}}\left( \begin{array}{c} v_j \\ 0 \end{array} \right)
~~~~{\rm and}~~~~
\langle \Psi_{jk} \rangle_0 ~=~
{1\over{\sqrt{2}}}
\left( \begin{array}{cc} v^{(jk)}_{1} & 0 \\ 0 & v^{(jk)}_{2}
\end{array} \right) \eeq
are the vacuum expectation values (VEVs) of the various doublets and
bidoublets and
\beq
\Phi_j'~\equiv~\Phi_j - \langle \Phi_j \rangle_0,~~~~~
\Psi_{jk}'~\equiv~\Psi_{jk} - \langle \Psi_{jk} \rangle_0.
\eeq
We choose to work in the Landau gauge $(\xi=0)$.
With respect to fermions, since we  assume this model to be
fermiophobic as in \cite{FP},
the unique gauge-invariant coupling that can be constructed between
fermions
and scalars involves only the $\Phi_{0}$ doublet, so the couplings are
the same as in the SM. A first positive consequence
is that one automatically avoids having FCNC problems at tree level,
in contrast to the usual left-right models. Also, no mass term is
introduced
for the right-handed neutrino.

Finally, the Lagrangian of our model can be obtained by combining these
various terms, i.e. 
\beq
\label{lagr}
{\cal L}~\equiv~{\cal L_{{\rm YM}}}~+~{\cal L_{{\rm sca}}}~+~
{\cal L_{{\rm gf}}}.
\eeq

\section { \bf  Gauge boson mass matrices}
 \bigskip

The kinetic term of the scalar Lagrangian in eq.(\ref{scalagr}) generates,
 after the
spontaneous symmetry breaking, a mass term for the charged
($W^{\pm}_{j}$ with $j=0,1,\cdots.n$) and neutral gauge bosons ($Z_{j}$
with $j=0,1,\cdots,n$).

\subsection{Charged sector}

The charged $W^{\pm}_{j}$ boson mass matrix is an $(n+1) \times (n+1)$
symmetric matrix. 
In a compact way it can be written as
\beq \label{massc}
M_{C\, ij}^{2}={g_{0}^{2} \over 4} x_{i-1} x_{j-1} \left(\delta_{ij} (
v_{i-1}^{2}+s(i) ) - \theta(j-i-1) 2 v_{1}^{(i-1 \,
j-1)} v_{2}^{(i-1 \, j-1)}\right) \quad {\rm with} {\quad i\le j}
\eeq 
where 
$\theta(i)$
stands for the Heaviside function (with $\theta(0)=1$), 
$u_{ij}^{2}=v^{(i\, j)
\,2}_{1}+v^{(i \,j) \,2}_{2}$ \footnote{ $u_{ij}$ is obviously
symmetric
under the interchange of its indices.} and
\beq s(k)={\displaystyle
\sum^{n}_{i=0 \, i\neq k-1}} u_{i\, k-1}^{2}.
\eeq
The mass eigenvalues
 can be  obtained
from eq.(\ref{massc}) by performing an orthogonal transformation

\beq  \label{eqd}
M_{C}^{2 \, d}=R_{C} M_{C}^{2} R_{C}^{T}
\eeq
where $M_{C}^{2 \, d}$ is the charged diagonal mass matrix and
$R_{C}$ is the rotation matrix that transforms the interacting charged
 $W^{\pm}_{j}$ fields into the physical mass eigenstates $W^{P \pm}_{j}$

\beq \label{rot1}
\left(\begin{array}[c]{c} W^{P \pm}_{0} \\ W^{P \pm}_{1} \\
W^{P \pm}_{2} \\ \vdots \\W^{P \pm}_{n}
\end{array} \right) = R_{C}
\left(\begin{array}[c]{c} W^{\pm}_{0} \\ W^{\pm}_{1} \\
W^{\pm}_{2} \\ \vdots \\W^{\pm}_{n}
\end{array} \right)
\eeq
This rotation matrix $R_{C}$ involves
${n (n+1) \over 2}$ mixing angles.
The specific form of the rotation matrix $R_{C}$ and the mass
eigenvalues will be given in section 5 for the screening
configurations.

\subsection{Neutral sector}

The neutral mass matrix in the
interacting basis ($W_{0}^{3},W_{1}^{3},W_{2}^{3},\ldots,W_{n}^{3},B$)
is an $(n+2)\times (n+2)$ matrix given by
\beq \label{massn}
M_{N\, ij}^{2}={g_{0}^{2} \over 4} x_{i-1} x_{j-1} \left(\delta_{ij} (
v_{i-1}^{2}+s(i) ) - \theta(j-i-1) u_{i-1 \,
j-1}^{2}\right) \quad {\rm with} {\quad i\le j}
\eeq
where, to keep the notation as compact as possible, we have defined
\beq
u_{n+1\, i}^{2}\equiv v_{i}^{2} \qquad {\rm and}   \qquad
v_{n+1}^{2}\equiv 0.
\eeq
The diagonalization of this mass matrix is done in two steps.
First, we perform a finite
rotation with a  unitary matrix $U$ to identify the photon
field. This transforms the matrix of eq.(\ref{massn}) in a block form
with the first row and column filled with zeros corresponding to the
zero mass eigenvalue of the photon.
Afterwards, a second rotation $R_{N}$
is performed to obtain the mass eigenstates of the $Z_{i}$ fields.

Accordingly, the transformations from the interacting ($W_{0}^{3},W_{1}^{3},
W_{2}^{3},\ldots,W_{n}^{3},B$) to
the mass eigenstates basis ($A,Z_{0}^{P},Z_{1}^{P},Z_{2}^{P},\ldots,
Z_{n}^{P}$) are
\bea \label{rot2}
\left(\begin{array}[c]{c} A \\ Z_{0} \\
Z_{1} \\ \vdots \\Z_{n-1} \\ Z_{n}
\end{array} \right) = U
\left(\begin{array}[c]{c} W^{3}_{0} \\ W^{3}_{1} \\
W^{3}_{2} \\ \vdots \\ W^{3}_{n} \\ B
\end{array} \right) \nonumber \quad {\rm and} \quad
\left(\begin{array}[c]{c} A \\ Z^{P}_{0} \\
Z^{P}_{1} \\ \vdots \\Z^{P}_{n-1} \\ Z^{P}_{n}
\end{array} \right) = R_{N}
\left(\begin{array}[c]{c} A \\ Z_{0} \\
Z_{1} \\ \vdots \\ Z_{n-1} \\ Z_{n}
\end{array} \right) \nonumber \quad
{\rm where } \quad
R_{N}~=~\left(
\begin{array}{ccccccc}
1       &\ldots   0  &\ldots          & 0\\
\vdots  &            &                &  \\
0       &            &{\tilde R}_{N} &  \\
\vdots  &            &                &  \\
0       &            &                &  \\
\end{array}
\right) \nonumber
\eea
The matrix ${\tilde R}_{N}$ has the same structure
as the matrix
$R_{C}$.

The finite matrix $U$ has been built in two steps. In the first step,
the first row has been constructed with the constraint of being orthogonal
to
the mass matrix of eq.(\ref{massn}).
In a second step, all the other rows have been constructed
to be orthogonal to the first one. A different choice of the orthogonal
rows simply implies a redefinition of the mixing angles of the matrix $R_{N}$.
The
resulting matrix $U$ is 
\beq \label{matuik}
U_{ik}=\delta_{i\, 1} {P(x_{k-1}) \over f(-1)} + \delta_{k \, i-1}
 {f(k-1) \over f(k-2)} - \theta(k-i) \theta(i-2) {P(x_{k-1})
P(x_{i-2}) \over f(i-2) f(i-3)}
\eeq
where  the functions $P(x_{i})$
and $f(i)$
are defined as
\beq
P(x_i) = {{x_1\cdots x_{n} y}\over{x_i}} \qquad {\rm and} \qquad
f(i)=\sqrt{{\displaystyle
\sum_{s=i+1}^{s=n+1}} P(x_{s})^{2}}
\eeq
Notice that $f(n+1)=0$.

This is a
generalization for arbitrary $n$ of the $U$ matrix introduced in
\cite{alt} for $n=1$. Applying the transformation
matrix $U$ to $M_{N}^{2}$
\beq {\tilde M}_{N}^{2}=U M_{N}^{2} U^{T}
\eeq
we obtain the following matrix elements:
 \bea \label{mijn}
{\tilde M}_{N\,ij}^{2}&=&{g_{0}^{2} \over 4} \theta(i-2) \theta(j-2) \Biggl\{
F_{1}^{ij} \left[ \delta_{ij} (v^{2}_{j-2}+s(j-1))
- \theta(j-i-1)  u^{2}_{j-2 \, i-2} \right]
 \nn \\
&+& F_{2}^{ij} \left[ \sum_{r=j}^{n+2} \left[
\left( \sum_{s=0}^{i-2}
u^{2}_{s \, r-1}\right) + {{f(i-2)}^{2} \over {P(x_{i-2})}^{2}}
u^{2}_{ r-1 \, i-2} +
 {{f(j-2)}^{2} \over {P(x_{j-2})}^{2}} u^{2}_{
r-1 \, j-2} \right] \right.  \\ &+&\left. \left.
{{f(j-2)}^2 \over {P(x_{j-2})}^{2}}\left[ \left(\sum_{r=i}^{
j-2} u^{2}_{r-1 \, j-2}\right) - \theta(j-(i+1)) (v^{2}_{j-2}+
s(j-1)) \right] \right] \right\} \quad {\rm with} {\quad i \leq j} \nn
\eea
with
\bea
F_{1}^{ij}&=&x_{i-2} x_{j-2} {f(j-2) f(i-2) \over f(j-3)
f(i-3)} \nn \\
F_{2}^{ij}&=&{{P(x_{0})}^4 \over x_{i-2} x_{j-2}}{1 \over f(i-2)
f(i-3) f(j-2) f(j-3)}.
\eea
The matrix elements of $M_{N}$ have been denoted by ${\tilde M}_{N}$ to
recall that they are not the final result, because  it is still necessary
to
make the second rotation
on the $Z$ fields
in order to get the mass eigenvalues:
\beq
M_{N}^{2 \, d}=R_{N} {\tilde M}_{N}^{2} R_{N}^{T}.
\eeq
The complete rotation matrix is denoted by

\beq \label{calr} {\cal R}=R_{N} \times U.\eeq
The precise form of the mass eigenvalues and
the explicit dependence on the VEVs and coupling constants of the rotation
matrix $R_{N}$ or ${\cal R}$ will be given in section 5 for the screening
configurations.

\section { \bf  Scalars}
 \bigskip

In this section we analyse the spectrum of scalar particles
of the theory, Goldstones and Higgses.
In the interacting basis of scalar fields we have
$4(n+1)$ degrees of freedom (d.o.f.) coming from the doublets
together with $4 n (n+1)$ d.o.f. from the bidoublets. Half of them
are charged, while the others are neutral.

\subsection{\bf Charged Goldstone bosons and Higgses}

The charged d.o.f. combine in
$2(n+1)$ charged Goldstone bosons $G_{i}^{\pm}$ ($i=0,\ldots,n$) and
$2 n (n+1)$  states  corresponding to the physical charged
Higsses $H_{i}^{\pm}$ ($i=n+1,\ldots,n(n+2)$).

We will start by identifying the Goldstone bosons
that will be ``eaten" up to give masses
to their corresponding gauge bosons.

From the Lagrangian in eq.(\ref{scalagr}), after performing the rotations
of the
gauge fields from the interacting to the mass eigenstates,
eq.(\ref{rot1}),
we identify each
Goldstone boson as the
coefficient of the term linear in the corresponding physical gauge boson.

Then we introduce a unitarity matrix ${\cal A}$
that gives the projection of the
interacting fields onto the mass eigenstates

\bea \label{mata}
\left(\begin{array}[c]{lll} &&\phi^{\pm}_{0} \\ &&\phi^{\pm}_{1} \\
&&\vdots \\ &&\phi^{\pm}_{n} \\ &&\varphi^{\pm \, (0 \, 1)}_{1} \\
&&\varphi^{\pm \, (0 \,1)}_{2} \\ &&\vdots
\\
&&\varphi^{\pm \, (j \, l)}_{1} \\
&&\varphi^{\pm \, (j \, l)}_{2} \\
&&\vdots  \\
 &&\varphi^{\pm \,(n-1 \,
n)}_{1} \\ &&\varphi^{\pm \, (n-1 \, n)}_{2}\end{array} \right)
\, = \, \, {\cal A} \,
\left(\begin{array}[c]{lll} &&G^{\pm}_{0} \\ &&G^{\pm}_{1} \\
&&\vdots \\ &&G^{\pm}_{n} \\ &&H^{\pm}_{n+1} \\
&&H^{\pm}_{n+2} \\ &&\vdots \\ &&\vdots  \\&&\vdots \\ &&\vdots \\&&\vdots
 \\ &&H^{\pm}_{n(n+2)}\end{array} \right)
\quad
C^{i}_{\cal A}~=~\left(\begin{array}[c]{lll}
&& {g_{0} (R_{C})_{i \,1} v_{0} \over N_{i}}
\\
&&{ g_{1} (R_{C})_{i \,2} v_{1} \over N_{i}}
\\
&& \vdots
 \\
&&{ g_{n} (R_{C})_{i \,n+1} v_{n} \over N_{i}}
\\
&&{g_{0}  (R_{C})_{i \, 1} v_{1}^{(0\,1)} - g_{1}
(R_{C})_{i \,2} v_{2}^{(0\,1)} \over N_{i}}
 \\
&&{-g_{0}  (R_{C})_{i\,1} v_{2}^{(0\,1)}
+ g_{1}
 (R_{C})_{i \,2} v_{1}^{(0\,1)} \over N_{i}}
 \\
&&\ldots  \\
&& {g_{j}  (R_{C})_{i\,j+1} v_{1}^{(j\, l)} - g_{l}
(R_{C})_{i\,l+1} v_{2}^{(j\, l)} \over N_{i}}
 \\
&&{-g_{j} (R_{C})_{i\, j+1} v_{2}^{(j\, l)}
 + g_{l} (R_{C})_{i\, l+1} v_{1}^{(j\, l)} \over N_{i}}
 \\
&&\vdots
  \\
&&{ g_{n-1} (R_{C})_{i\,n} v_{1}^{(n-1\, n)} - g_{n}
(R_{C})_{i\,n+1}  v_{2}^{(n-1\, n)} \over N_{i}}
 \\
&&{ -g_{n-1} (R_{C})_{i\,n}  v_{2}^{(n-1\, n)} + g_{n}
(R_{C})_{i\,n+1}  v_{1}^{(n-1\, n)} \over N_{i}}
\end{array} \right)
\eea
where the normalization factor $N_{i}$ is given by
\bea
N_{i}=g_{0}\sqrt{{\displaystyle \sum_{s=0}^{s=n}} x_{s}^{2} v_{s}^{2}
{(R_{C})^{2}_{i\,s+1}}+
{\displaystyle \sum_{j=0}^{n-1} \sum_{l=j+1}^{n}}
\left[
{\left( x_{j} (R_{C})_{i \, j+1} v_{1}^{(j\,l)}-x_{l} (R_{C})_{i \, l+1}
v_{2}^{(j \, l)} \right)}^{2}+v_{1} \leftrightarrow v_{2}
 \right]  }.
\eea
The rotation matrix ${\cal A}$ can be split into two sets of columns
${\cal A}=\left\{ C_{\cal A}^ {i} , C_{\cal A}^{ \alpha} \right\}$.
The first set  $C^{i}_{\cal A}$ (given in eq.(\ref{mata})),  with $i$
running
from $1$ to $n+1$,
corresponds precisely to the decomposition of the charged Goldstone boson
$G_{i-1}^{\pm}$ with mass $m_{i-1}^{\cal A}$ in terms of the interacting
basis.
The second set $C^{\alpha}_{\cal A}$ with $\alpha=n+2,\ldots,(n+1)^{2}$
is the corresponding projection of the mass eigenstate $H_{\alpha-1}^{\pm}$
\footnote{Notice that we use a different index notation for the
$H_{i}^{\pm}$ and their mass than in \cite{FP}. Our first physical
charged Higgs is named $H_{n+1}^{\pm}$ instead of  $H_{1}^{\pm}$ of
\cite{FP}, to be
able to get a compact  expression for $  \rho_{\rm NC}$ in the next
section.}
whose mass is denoted by $m_{\alpha-1}^{\cal A}$. We do not give here the
explicit form of   $C^{\alpha}_{\cal A}$ since we
do not need it to find the screening configurations. In order to construct
$C^{\alpha}_{\cal A}$ it
is just necessary to find a set of states orthonormal among
themselves and
to the Goldstone bosons. For an explicit example in the case $n=1$, see
\cite{FP}. Notice
that the choice of this
basis of orthonormal states leaves  extra freedom to
introduce $n(n+1)(n(n+1)-1)/2$ angles.

\subsection{Neutral Goldstone bosons and Higgses}

The neutral sector is organized in a set of  $(n+1)$ neutral Goldstone bosons
and $(2n+1)(n+1)$ neutral physical Higgses. We assume here, as in
\cite{FP}, that there are no other sources of CP-violation apart from
the
phase of the Kobayashi-Maskawa matrix. This hypothesis
allows us to split the neutral
scalars in a  CP-odd set ($n(n+1)$ neutral physical Higgses $H_{i}^{0}$
$i=n+1,\ldots,n(n+2)$ and  $n+1$
Goldstone bosons $G_{i}^{0}$ $i=0,\ldots, n$) and a CP-even set (${(n+1)}^{2}$
neutral Higsses) without mixing between the two sets.

In a similar way to the charged case, the rotation of the neutral gauge
bosons from the interacting to the mass eigenstate basis in the Lagrangian
of eq.(\ref{scalagr}), using eq.(\ref{calr}), provides us with the
definition of the neutral Goldstone bosons in terms of the interacting
fields.

Also here we introduce  a matrix for the physical CP-odd states and the
neutral would-be Goldstone bosons (called ${\cal
C}$) 
that gives the projection of the
interacting fields onto the mass eigenstates basis:
\bea \label{matc}
\left(\begin{array}[c]{lll} &&{\rm Im}\phi^{0}_{0} \\
&&{\rm Im}\phi^{0}_{1}
\\ &&\vdots \\ &&{\rm Im}\phi^{0}_{n} \\ &&{\rm Im}\varphi^{0 \, (0 \,
1)}_{1} \\ &&{\rm Im}\varphi^{0 \, (0 \,1)}_{2} \\ &&\vdots
\\
&&{\rm Im}\varphi^{0 \, (j \, l)}_{1} \\
&&{\rm Im}\varphi^{0 \, (j \, l)}_{2} \\
&&\vdots  \\
 &&{\rm Im}\varphi^{0 \,(n-1 \,
n)}_{1} \\ &&{\rm Im}\varphi^{0 \, (n-1 \, n)}_{2}\end{array} \right)
\, = \, \, {\cal C} \,
\left(\begin{array}[c]{lll} &&G^{0}_{0} \\ &&G^{0}_{1} \\
&&\vdots \\ &&G^{0}_{n} \\ &&H^{0}_{n+1} \\
&&H^{0}_{n+2} \\ &&\vdots \\ &&\vdots  \\&&\vdots \\ &&\vdots \\&&\vdots
 \\ &&H^{0}_{n(n+2)}\end{array} \right)
\qquad
C^{i}_{\cal C}~=~\left(\begin{array}[c]{lll}
&&{ \left(g_{0} {\cal R}_{i+1 \,1}
-{\tilde g} {\cal R}_{i+1 \, n+2}\right) v_{0} \over {\tilde N}_{i}}
\\
&&{\left( g_{1}  {\cal R}_{i+1 \,2}
-{\tilde g} {\cal R}_{i+1 \, n+2}\right) v_{1} \over {\tilde N}_{i}}
\\
&&\ldots
\\
&& {\left(g_{n} {\cal R}_{i+1 \,n+1}
-{\tilde g} {\cal R}_{i+1 \, n+2}\right) v_{n} \over {\tilde N}_{i}}
\\
&&{\left(g_{0}  {\cal R}_{i+1 \, 1} - g_{1} {\cal R}_{i+1 \,2} \right)
v_{1}^{(0\,1)}
\over {\tilde N}_{i}}
\\
&&{-\left(g_{0} {\cal R}_{i+1 \,1} - g_{1}  {\cal R}_{i+1 \,2} \right)
v_{2}^{(0\,1)}
\over{\tilde N}_{i}}
\\
&&\vdots
\\
&&{\left(g_{j} {\cal R}_{i+1 \,j+1} - g_{l} {\cal R}_{i+1 \,l+1} \right)
v_{1}^{(j\,l)}
\over
{\tilde N}_{i}}
\\
&&{-\left(g_{j} {\cal R}_{i+1 \, j+1} - g_{l} {\cal R}_{i+1 \, l+1} \right)
v_{2}^{(j\,l)}
 \over {\tilde N}_{i}}
\\
&&\vdots
\\
&& {\left(g_{n-1} {\cal R}_{i+1 \,n} - g_{n} {\cal R}_{i+1 \,n+1} \right)
v_{1}^{(n-1\, n)}
\over {\tilde N}_{i}}
\\
&&{ -\left(g_{n-1}  {\cal R}_{i+1 \, n} - g_{n}
{\cal R}_{i+1 \,n+1} \right)
v_{2}^{(n-1\, n)}
\over {\tilde N}_{i}}
\end{array}
\right)
\eea
where
\beq \label{t}
{\cal R}_{ij}=\delta_{i \,1} {P(x_{j-1}) \over f(-1)}
+ {({R}_{N})}_{i\, j+1} {f(j-1)
\over f(j-2)} - {\displaystyle \sum_{s=2}^{n+2}} \theta(j-s)
{({R}_{N})}_{i \, s}  {P(x_{j-1}) P(x_{s-2}) \over f(s-2)
f(s-3)}
 \eeq
with $i,j=1,...,n+2$ and the normalization factor is
\beq
{\tilde N}_{i}=g_{0}\sqrt{{\displaystyle \sum_{k=0}^{k=n}}{(x_{k}
{\cal R} _{i+1
\, k+1}-y {\cal R}_{i+1 \,n+2})}^{2} v_{k}^{2} +
{\displaystyle \sum_{j=0}^{n-1} \sum_{l=j+1}^{n}}{(x_{j} {\cal R}_{i+1 \,
j+1}-x_{l} {\cal R}_{i+1 \, l+1})}^{2} u_{jl}^{2}}.
\eeq
We can split the matrix into two sets of columns
${\cal C}=\left\{ C_{\cal C}^ {i} , C_{\cal C}^{ \alpha} \right\}$,
with the same conventions
for the column indices as in the charged case.
The column $C_{\cal C}^{i}$ given in
eq.(\ref{matc})
corresponds to the neutral Goldstone boson $G^{0}_{i-1}$ with mass $m^{\cal
C}_{i-1}~~~(i=1,\dots,n+1)$. 
The column index $\alpha$ runs from $n+2$ to $(n+1)^{2}$ and labels the
set of
physical CP-odd neutral Higgses $H_{\alpha-1}^{0}$ with mass
$m_{\alpha-1}^{\cal C}$. The construction of the columns $C^{\alpha}_{\cal C}$
follows the same rules as in the charged case.
Also in this case one has the
freedom to define $n(n+1)(n(n+1)-1)/2$ angles.

There is also a third matrix, which we have called ${\cal B}$,
that relates
the real part of the neutral fields with a set of CP-even $(n+1)^{2}$ physical
Higgses with mass $m_{i-1}^{\cal B}$ $(i=1,\ldots,{(n+1)}^{2})$. We do
not give an explicit form for this matrix since in
the screening configurations it is always taken to be equal to ${\cal
A}$ and ${\cal C}$, except for the rows concerning the fields
$\varphi_{2}^{0 \, (jl)}$ whose sign in matrix $\cal B$ is reversed with
respect to the other matrices due to the conventions used in the
bidoublets sector. This requirement simply fixes the rotation angles
in the matrix ${\cal B}$, and we always have enough
freedom to choose them. The first column $C^{1}_{\cal B}$ corresponds to
the SM
Higgs-like with a mass $m_{0}^{\cal B}$.

\bigskip
\section { \bf  Computation of $  \rho_{\rm NC}$ }
 \bigskip

In this section we give the explicit expression of
the leading contribution (quadratic in the scalar masses) to
$\Delta \rho_{\rm NC}\equiv\rho_{\rm NC}-1$ for arbitrary $n$.

In a renormalizable theory, at one loop, the contribution to
$\Delta\rho_{\rm NC}$ can grow, by power counting, at most quadratically.
As we have said before, we are interested in models for which
there are natural configurations
of the VEVs and of the masses of the scalars such that
the potentially large $m^{2}$ contributions to $\Delta \rho_{\rm NC}$
cancel and
we have a screening phenomenon.

The $\rho_{\rm NC}$ parameter
is defined following \cite{sir}
as the ratio between
the neutral and the charged current couplings, and therefore the
expression of $\Delta\rho_{\rm NC}$, evaluated at $q^2=0$, is:
\beq
\label{rhodef}
\Delta \rho_{\rm NC}={\Sigma_{W}(0) \over M_{W}^{2}}-{\Sigma_{Z}(0)
\over M_{Z}^{2}}
\eeq
where  $\Sigma_{V}(q^{2})$ (with $V=Z,
W$) stands for the
transverse part of the self-energy corrections to the propagator of the
vector boson $V$. The contribution coming from $\Sigma_{\gamma Z}(0)$ 
is absent since we have checked that it is exactly zero by the
orthogonality properties of matrix ${\cal A}$.

We use this definition of $\rho_{\rm NC}$, at $q^2=0$, since we are
interested
in corrections that are leading, universal and independent of the
value of $q^2$.
The kind of diagrams that can give a 
quadratic contribution, in the masses of
the scalar fields,
to eq.(\ref{rhodef})
 are shown in Fig.1.
Their explicit expression is:
\begin{eqnarray*}
{z[m_{i}]} &=&-{m_{i}^{2} \over  {16\pi^{2}}}\left(
C_{UV}+1-\ln m_{i}^{2}\right)  \\
{t[m_{i},m_{j}]} &=&{1 \over 16\pi ^{2}}\left[ \left(
C_{UV}+3/2\right) \left( m_{i}^{2}+m_{j}^{2}\right) -\left( \frac{
m_{i}^{4}
\ln m_{i}^{2}-m_{j}^{4}\ln m_{j}^{2}}{m_{i}^{2}-m_{j}^{2}}\right)
\right]
\end{eqnarray*}
where $C_{UV}=1/\epsilon-\gamma+{\rm Log} 4 \pi$ and any symmetry factor
or
coupling constant has been factorized
out.

\begin{figure}
  \centering
\epsfig{file=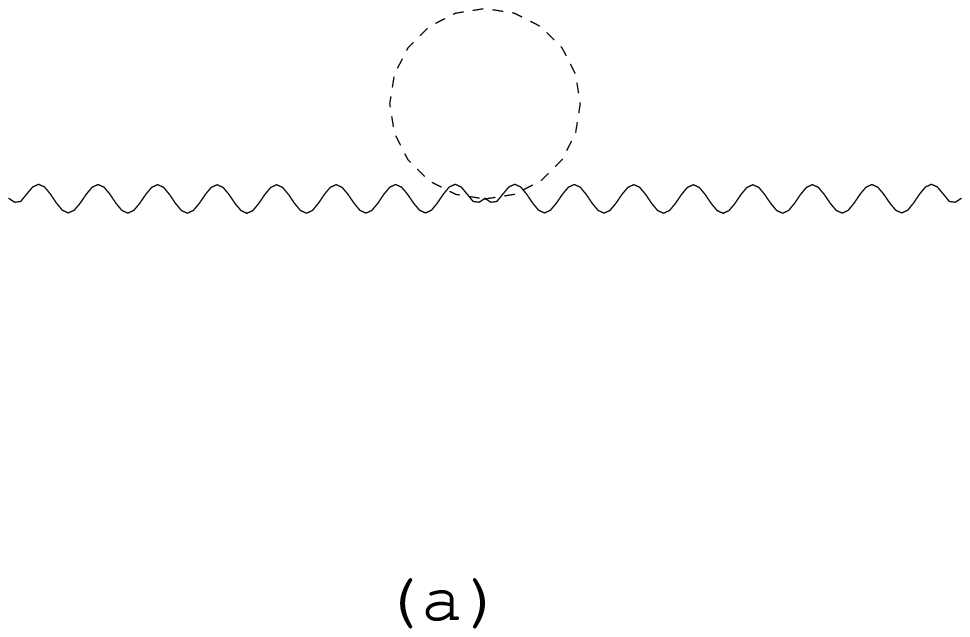,height=6cm,width=6cm,
        bbllx=0pt,bblly=0pt,bburx=300pt,bbury=300pt}
\epsfig{file=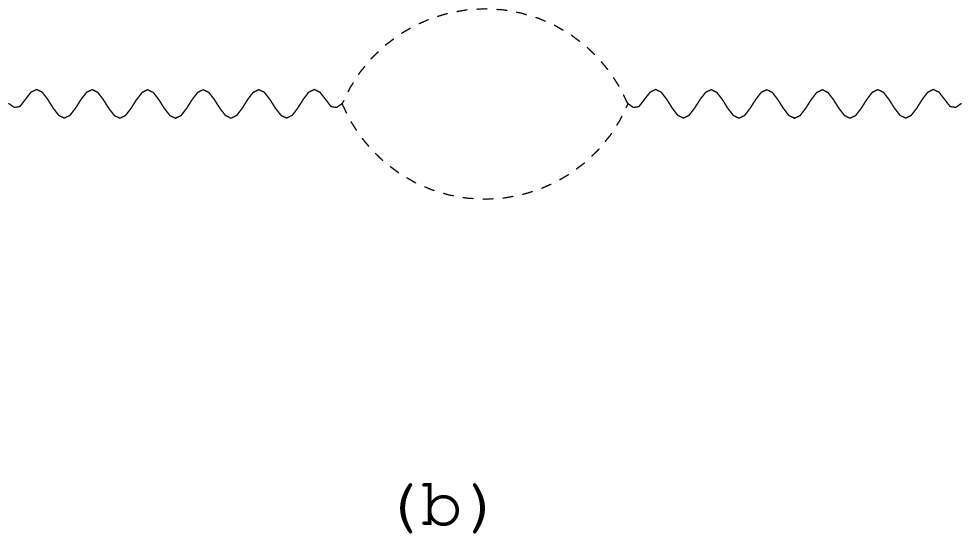,height=6cm,width=6cm,
        bbllx=-100pt,bblly=0pt,bburx=200pt,bbury=300pt}

\caption{Feynman diagrams contributing to $  \rho_{\rm NC}$, a)
$z[m_{i}]$ and b) $t[m_{i},m_{j}]$. The tadpole diagram (not included)
coming from the scalar potential only renormalizes the VEVs
and is not relevant for our computation \cite{vel,ew}.}
\end{figure}

If one expands the Lagrangian in eq.(\ref{scalagr}), keeping the vertices
involving
two gauge bosons ($W$, $Z$ and $A$) and two scalars and those involving
one gauge and two scalars, the result is

\bea \label{delta}
\Delta \rho_{\rm NC} &=&{1 \over 4 M_{W}^{2}}\left\{
 \sum_{l,m=0}^{n(n+2)} \left[ \left( \sum_{j=0}^{n}
{\cal A}_{j+1 \, l+1} {\cal B}_{j+1 \, m+1} P_{j}+
\sum_{j=0}^{n-1} \sum_{k=j+1}^{n}
\biggl[ \left( {\cal A}_{s1 \, l+1} {\cal B}_{s1 \, m+1}
-{\cal A}_{s2 \, l+1} {\cal B}_{s2 \, m+1}\right) P_{j} \right. \right.
\right. 
\nn \\
&+&\left. \left. \left. \left. \left({\cal A}_{s2 \, l+1} {\cal B}_{s1
\, m+1}
-{\cal A}_{s1 \, l+1} {\cal B}_{s2 \, m+1} \right) P_{k} \biggr] \right)^{2}
f[m_{l}^{\cal A},m_{m}^{\cal B}] \right] \right.  \right.
\nn \\
&+&  \sum_{l,m=0}^{n(n+2)} \left[ \left( \sum_{j=0}^{n}
{\cal A}_{j+1 \, l+1} {\cal C}_{j+1 \, m+1} P_{j}+
\sum_{j=0}^{n-1} \sum_{k=j+1}^{n}
\biggl[ \sum_{f=1}^{2} {\cal A}_{s_{f} \, l+1} {\cal C}_{s_{f} \, m+1}
 P_{j} \right. \right.
 \nn \\
&-&\left. \left. \left. \left. \left({\cal A}_{s2 \, l+1} {\cal C}_{s1
\, m+1}
+{\cal A}_{s1 \, l+1} {\cal C}_{s2 \, m+1} \right) P_{k} \biggr] \right)^{2}
f[m_{l}^{\cal A},m_{m}^{\cal C}] \right] \right\}  \right.
\nn \\
&-&{1 \over 4 M_{Z}^{2}}\left\{ 
\sum_{l,m=0}^{n(n+2)} \left[ {\left( \sum_{j=0}^{n}
{\cal A}_{j+1 \, l+1} {\cal A}_{j+1 \, m+1} N_{1\,j}^{+}+
\sum_{j=0}^{n-1} \sum_{k=j+1}^{n} \sum_{f=1}^{2}
 {\cal A}_{s_{f} \, l+1} {\cal A}_{s_{f} \, m+1}
  N_{2\, jk}^{+} \right) }^{2}
f[m_{l}^{\cal A},m_{m}^{\cal A}] \right. \right.
\nn \\
&+&\left. \left.   {\left( \sum_{j=0}^{n}
{\cal C}_{j+1 \, l+1} {\cal B}_{j+1 \, m+1} N_{1 \,j}^{-}+
\sum_{j=0}^{n-1} \sum_{k=j+1}^{n}
\left( {\cal C}_{s1 \, l+1} {\cal B}_{s1 \, m+1}
- {\cal C}_{s2 \, l+1} {\cal B}_{s2 \, m+1} \right) N_{2 \,jk}^{-} \right) }^{2}
f[m_{m}^{\cal B},m_{l}^{\cal C}] \right] \right\}, \nn \\ &&
\eea
where we have used that
\beq \label{decomp}
t[m_{i},m_{j}]=-z[m_{i}]-z[m_{j}]+f[m_{i},m_{j}]
\eeq
with
\beq
f[m_{i},m_{j}]={1 \over 16 \pi^{2}}\left[{1 \over 2}(m_{i}^{2}+m_{j}^{2})
-{m_{i}^{2} m_{j}^{2} \over m_{i}^{2}-m_{j}^{2}} {\rm Log}[
{m_{i}^{2} \over m_{j}^{2}}] \right].
\eeq
We have introduced, for convenience, the indices $s_f~~(f=1,2)$:
\beq
s_1=n+2 \left[ \left(n-\left({j+1 \over 2} \right)\right) j + k \right]
\qquad \qquad \qquad
s_2=s_1+1;
\eeq
$s_1$ labels the fields $\phi_1^{\pm~(jk)}$,
${\rm Re} \phi_1^{0~(jk)}$
 and
${\rm Im} \phi_1^{0~(jk)}$, while $s_2$ 
does the same for $\phi_2^{\pm~(jk)}$,
${\rm Re} \phi_2^{0~(jk)}$ 
 and
${\rm Im} \phi_2^{0~(jk)}$.
The functions $P,N_{1}^{\pm}$ and $N_{2}^{\pm}$ are defined as
\bea
P_{x}= g_{x} \, (R_C)_{1 \, x+1},
\qquad
N_{1 \, j}^{\pm}=g_{j} {\cal R}_{2 \, j+1} \pm {\tilde g} {\cal R}_{2 \,
n+2},
\qquad
N_{2 \, jk}^{\pm}=g_{j} {\cal R}_{2 \, j+1} \pm g_{k} {\cal R}_{2 \, k+1}.
\eea
All the masses of the Goldstone bosons $(m_i^{{\cal A}}~{\rm and}~
m_i^{{\cal C}},~i=0,1,\dots,n)$ should be taken equal to zero, since we are
working in the Landau gauge.

The expression of $\Delta \rho_{\rm NC}$ remarkably simplifies after the
use of  eq.({\ref{decomp}) and of the ortogonality properties of the
matrices ${\cal A}$, ${\cal B}$
and ${\cal C}$, independently of their explicit expression.
From eq.(\ref{delta}) we notice that all the divergences cancel, indeed
they cancel separately inside
each self-energy; furthermore also all the finite contributions from
$z[m_i]$ cancel.
The function $f[m_i,m_j]$ gives a measure of the isospin breaking
and its properties serve as a guideline to find the
conditions to obtain the screening.
In fact $f[m_i,m_i]=0$ helps to find the conditions concerning the
degeneracy of the scalar masses, while $f[m_i,0]=m_i^2/(32 \pi^2)$  
gives us the constraints on the VEVs in order to compensate the scalar
contributions between the two self-energies.
The particular conditions will be discussed in the next section.

Also, we have verified that the previous expression reduces
for the case $n=1$ to the one of \cite{FP}, in all the cases that
are explicitly given there.

From a direct inspection of eq.(\ref{delta}), one notices the absence of
terms
proportional to $f[m_i^{{\cal B}}, m_j^{{\cal B}}]$ and to
$f[m_i^{{\cal C}}, m_j^{{\cal C}}]$, i.e. the absence of self-energy
diagrams with two neutral scalars or two neutral physical pseudo-scalars
running in the loop.
On the other side, in the unitary gauge, the self-energy diagrams
involving
one physical scalar and one Goldstone boson are obviously absent and their
contribution is shifted to the diagrams with the Goldstone boson replaced
by the gauge field.
From the previous two observations we learn that the contribution from
the diagrams in Fig.1 vanishes in the unitary gauge, if the scalar sector
contains either only physical scalars or only physical pseudoscalars
(cf. e.g. \cite{CL}).

\bigskip
\section { \bf  Screening Configurations}
 \bigskip

It was observed in \cite{FP} for the case $n=1$ that
under certain conditions the contributions to $\rho_{\rm NC}$, which are 
quadratic
in
the masses of the scalar particles, vanish.

We have found that  also when the gauge group is enlarged to
$SU(2)_{L} \times SU(2)^{n} \times U(1)_{Y}$, there are
at least two configurations in which these contributions to
$\rho_{\rm NC}$ vanish. 
The requirements imposed on the VEVs and consequently the
predicted mass spectrum of the gauge bosons are 
one of the differences between the configurations.
In the first one, which we will refer to as case I, the masses of all the
new gauge bosons are sent to infinity, while in the second one, referred
to as
case II, a finite-mass
spectrum for the new gauge particles is allowed.

A common condition to both configurations is the requirement of 
alignment of the scalar transformation
matrices ${\cal A}, {\cal B}$ and ${\cal C}$ or, to be more precise:

\beq \label{cond}
{\cal A}_{\alpha \, \beta}={\cal B}_{\alpha \, \beta}={\cal
C}_{\alpha \, \beta} \qquad {\rm except \quad  for} \qquad {\cal B}_{s2 \,
\beta}=-{\cal A}_{s2 \, \beta}=-{\cal C}_{s2 \, \beta}.
\eeq
The difference in sign is just due to the convention chosen to write
the $\phi_{2}^{(jk)}$ field inside the bidoublet $\psi_{jk}$.

In order to understand the implications of this first condition
on the parameter space, 
we should count the number of free parameters of
the gauge and scalar sector (without the potential).
We have on  one side $(n+1)^2$ VEVs and $n+2$ coupling
constants and, on the other,
 $n(n+1)(n(n+1)-1)/2$ mixing angles among the charged scalars. In the
neutral
sector, we have
 $n(n+1)(n(n+1)-1)/2$ and
$n(n+2)(n+1)^2/2$ mixing angles among  the neutral pseudoscalar and
the neutral scalar bosons, respectively.
Of course, we also have the set of masses of the physical Higgses,
already discussed.

This condition eq.(\ref{cond}) imposes constraints on the VEVs
and requires each mixing angle of matrix ${\cal A}$ to be equal
to the corresponding mixing angle in matrix ${\cal B}$ and ${\cal C}$.
So, at the end, in total, only $n(n+1)(n(n+1)-1)/2$ free
angles are left. Notice that the remaining mixing angles between the
neutral scalars are completely determined by eq.(\ref{cond}).

A second condition will be required on the masses of the physical
scalars.

The particular requirements on the VEVs and on the scalar masses
depend on the
case and will be discussed in the next subsections.

\bigskip
\subsection { \bf  Case I}
 \bigskip

In this first scenario, the set of equations given
by (\ref{cond}) are satisfied by taking the large-$v_{i}$ limit (with
$i>0$)
while keeping $v_{0}$, $v_{1}^{(i-1 \, j-1)}$ and
$v_{2}^{(i-1 \, j-1)}$ fixed and small with respect to the $v_{i}$'s.
When a hierarchy between the $v_{i}$'s is introduced
\beq \label{hie1}
v_{0}<v_{1}<v_{2}<...<v_{n}
\eeq
we find  the
expected alignment of the matrices ${\cal A}$, ${\cal B}$ and ${\cal C}$.
In that limit the leading-$v_{i}$ dependence of the charged and neutral
rotation matrices is, respectively,
\beq
{(R_{C})}_{i \, j} \sim {1 \over v_{j-1}^{2}} \qquad {\rm and} \qquad
{(R_{N})}_{i \,
j} \sim \theta(i-2) \theta(j-2)
{v_{i-2}^2 \over v_{j-2}^{2}},
\quad \quad i < j,
\eeq
where the elements along the diagonal are equal to 1 and the
off-diagonal ones change sign when $i>j$. The leading term of the
normalization factors is
\bea \label{lim}
N_{i}&\sim&g_{0} \sqrt{\delta_{i 1 }\left[v^{2}_{0}+s(1)\right]+
\theta(i-2) x^{2}_{i-1} v^{2}_{i-1}} \nn \\
{\tilde N}_{i}&\sim&g_{0}\sqrt{\delta_{i 1}{f(-1)^{2}
 \over f(0)^{2} } \left[v_{0}^{2}+s(1) \right]
+ \theta(i-2) x^{2}_{i-1} v^{2}_{i-1}{f(i-2)^{2}
\over f(i-1)^{2}}
}.
\eea
Now it is easy to see, by looking at the charged matrix ${\cal A}$
and at eq.(\ref{lim}), that in the large $v_{k}$ limit ($k=1,...$)
the only non-vanishing term of the column $i$ with $i>1$ is
the element in row $i$, which is equal to 1. All the other terms
are suppressed either by a mixing angle or
by the normalization factor.
The first column $i=1$ is exceptional since its $v_{0}$ is kept finite
and the normalization factor does not grow. In that case all the terms
proportional to a mixing angle are still vanishing while the others
remain. Then this column turns out to be

\beq
C^{1 \, T}_{\cal A}={1 \over \sqrt{v_{0}^{2}+s(1)}} \left({ v_{0}
, 0,...,0,\overbrace{
v_{1}^{(0 \, 1)}}^{\rm n+2 \; row}, - v_{2}^{(0 \,
1)},...,v_{1}^{(0 \, n)}, - v_{2}^{(0 \, n)},0,...,0}\right).
\eeq
It is not difficult to check, using eqs.(\ref{matc}) and (\ref{lim}),
that the same holds exactly for the neutral sector.
Notice that it is precisely the hierarchy of eq.(\ref{hie1}) that
guarantees that all the terms with a mixing angle vanish, since the VEVs in
the numerator cannot be larger than the ones in the denominator.
The standard gauge boson masses are in this case
\beq \label{casImas}
M_{W_{0}^P}^{2}={g_{0}^2 \over 4} (v_{0}^{2} + s(1)),~~~~ \quad
M_{Z_{0}^P}^{2}={g_{0}^2 \over 4} {f(-1)^{2} \over f(0)^{2}}
(v_{0}^{2}+s(1)),
 \quad
\eeq
where we have used that $f(0)^{2}=f(-1)^{2}-P(x_{0})^{2}$.
The new gauge boson masses are, at first order in the large $v_{i}$
expansion:       
\bea
M_{W_{i}^P}^{2}&=&{g_{0}^{2} \over 4} x_{i}^{2} \left( v_{i}^{2}+s(i+1)
\right)  \\
M_{Z_{i}^P}^{2}&=&{g_{0}^{2} \over 4} x_{i}^{2} {f(i-1)^2 \over f(i)^2}
\left\{v_{i}^{2}+s(i+1)+{P(x_{i})^4 \over f(i-1)^4} \sum_{s=0}^{s=i-1}
\left( v_{s}^{2} + \sum_{r=i+1}^{s=n} u_{s r}^{2} - \left({f(i)^{2}
+f(i-1)^2 \over P(x_{i})^2} \right) u_{is}^{2} \right) \right\}
\nn
\eea
and go to infinity when the large $v_i,~~i>1$ limit is taken.
 From the request that the coefficient in front of the
electromagnetic current be the electric charge,
one finds, once the matrix $U$ has been applied on the interacting fields,
that \beq \label{charge}
e=g_{0} P(x_{0})/f(-1).
\eeq
Using eq.(\ref{charge}) and the relation  $e=g s_{W}$, the sine and
cosine of the weak mixing angle are respectively given by
\beq \label{weinb}
s_{W}=P(x_{0})/f(-1) \quad {\rm and} \quad
c_{W}=f(0)/f(-1).
\eeq
Therefore, in this first configuration, it follows from
eqs.(\ref{casImas}) and (\ref{weinb}) that $\rho$
defined as
\beq \label{altrerho}
\rho=M_{W}^{2}/c_{W}^{2} M_{Z}^{2}
\eeq
is exactly 1.

This configuration gives us the possibility to
interpret in a simple way the alignment condition of eq.(\ref{cond}).
If one writes the Lagrangian of the model in this particular
configuration, by using matrices ${\cal A}$, ${\cal B}$ and ${\cal C}$,
which now turn out to be equal and very simple, one sees that the
Goldstones
and physical scalar fields are organized in a transparent way:

\beq
{\cal L}={\displaystyle \sum_{i=0}^{i=n}} {\left(D_{\mu} \Upsilon_{i}
\right)}^{\dagger} D_{\mu} \Upsilon_{i}+
{\displaystyle \sum_{i=1}^{i=n(n+1)}} {\left(D_{\mu} \Omega_{i}
\right)}^{\dagger} D_{\mu} \Omega_{i}
\eeq
where
\beq
\Upsilon_{i}=
\left(
\begin{array}{c}
G_{i}^{0}+i h_{i}^{0} \\
G_{i}^{-}
\end{array}
\right) \qquad {\rm and} \qquad
\Omega_{i}=
\left(
\begin{array}{c}
H_{i}^{0}+i h_{i+n+1}^{0} \\
H_{i}^{-}
\end{array}
\right).
\eeq
The set of $\Upsilon_{i}$ are complex doublets made of Goldstone
bosons and a singlet Higgs, and $\Omega_{i}$ are complex doublets of
matter. Thanks to the alignment, $\Delta \rho_{\rm NC}$
is nothing else than the sum of the contributions of a set of complex
doublets \cite{RK}. Therefore it, automatically, tells us which
are the conditions on the scalar masses to obtain the screening configuration:
\bea \label{split}
m_{l}^{\cal A}=m_{l}^{\cal B}  \qquad {\rm or}  \qquad
m_{l^{\prime}}^{\cal A}&=&m_{l^{\prime}}^{\cal C}
\eea
where $l,l^{\prime}=n+1,...,n(n+2)$.
We can now resume the main points of case I.
In order to get the alignment condition eq.(\ref{cond}), we perform the
large $v_i~~(i>0)$         
limit, with a hierarchy among the VEVs, and we set equal each
angle of matrix  ${\cal A}$ to the corresponding of matrix ${\cal B}$ and
${\cal C}$ (so at the end only $n(n+1)(n(n+1)-1)/2$ common angles are
left free).
 Then we see that the only remaining source of scalar quadratic terms is
given by 
the isospin breaking, i.e. the mass splitting in the doublets of 
eq.(\ref{split}). Once eq.(\ref{split}) is fulfilled we have screening.

\bigskip
\subsection { \bf  Case II}
\bigskip

This is the most interesting screening configuration. In fact a finite
light mass spectrum is allowed for the gauge bosons.
The result is obtained exactly, without the need of any approximation or
limiting procedure.
In this second scenario all the VEVs are kept finite. Obviously this
makes it much more involved to find the solution of eq.(\ref{cond}), so we
will give some details of the way to proceed.
We split the set of
equations ${\cal A}={\cal C}$ into two sets.
A first set is
\beq \label{mc1}
{x_{k-1} {\cal R}_{i+1 \, k}- {y} {\cal R}_{i+1 \, n+2} \over
x_{k-1} (R_{C})_{i \, k}}={{\tilde N}_{i} \over N_{i}},
\eeq
with $k=1,..,n+1$, while the second is
\bea \label{mc2}
{x_{j} {\cal R}_{i+1\,j+1} - x_{l} {\cal R}_{i+1\,l+1}
\over
x_{j} v_{1}^{(j\, l)} (R_{C})_{i\,j+1} - x_{l} v_{2}^{(j\,l)}
(R_{C})_{i\,l+1}
} v_{1}^{(j\,l)}
&=&{{\tilde N}_{i} \over N_{i}}
\nn \\
{x_{j} {\cal R}_{i+1\,j+1} - x_{l} {\cal R}_{i+1\,l+1}
\over
x_{j} v_{2}^{(j\, l)} (R_{C})_{i\,j+1} - x_{l} v_{1}^{(j\,l)}
(R_{C})_{i\,l+1}
} v_{2}^{(j\,l)}
&=&{{\tilde N}_{i} \over N_{i}}
\eea
 From the second set of equations one immediately obtains the
condition \beq
v_{1}^{(j \,l)}=v_{2}^{(j \,l)}.
\eeq
Notice that our VEVs are taken to be real.
Moreover, once this condition is imposed the fulfilment of
eq.(\ref{mc2}) is automatically ensured by the fulfilment of eq.(\ref{mc1}).
 This is easily seen by making the difference between
eq.(\ref{mc1}), evaluated at $k=j+1$, and the same equation evaluated at
$k=l+1$. It means that we can concentrate just on eq.(\ref{mc1}).
We write  eq.(\ref{mc1}) in the following way:
\beq \label{mc1b}
{x_{k-1} (R_{C})_{i \, k} \over
x_{k-1} {\cal R}_{i+1 \, k}- {y} {\cal R}_{i+1 \, n+2}}=
{x_{l-1} (R_{C})_{i \, l} \over
x_{l-1} {\cal R}_{i+1 \, l}- {y} {\cal R}_{i+1 \, n+2}} \quad {\rm with}
\quad k<l.
\eeq
 From the solution of the previous equation, we obtain  the constraints
on the VEVs,
\beq \label{vevcon}
v_{i}={v_{0} \over x_{i}} \qquad {\rm and} \qquad
u_{i \, j}^{2}=u_{i \, i+1}^{2} {\left({x_{i+1} \over x_{j}}
\right)}^{2},  \, \, \qquad j>i.
\eeq

The {\bf exact} form of the rotation matrices (charged and
neutral) corresponding to this screening configuration  is
the cornerstone of the calculation.
The charged matrix becomes

\bea \label{crot}
{(R_{C})}_{i \, k}=N^{C}_{i} \left( {\theta(k-i) \over x_{k-1}} -
\theta(i-2) \delta_{k \,i-1} x_{i-2} \sum_{j=k}^{n} {1 \over 
x_{j}^{2}} \right),
\eea
where
\beq
N^{C}_{i}={1 \over \sqrt{{\displaystyle \sum_{k=i-1}^{n}
{1 \over x_{k}^{2}}}\left(1+
\theta(i-2) x_{i-2}^{2}{\displaystyle \sum_{j=i-1}^{n}
{1 \over x_{j}^{2}}} \right)}},
\eeq
while the neutral one is

\beq \label{rnet}
{\cal R}_{i \, k}=\delta_{i \, 1}{P(x_{k-1})
 \over f(-1)} + \theta(i-2) S_{i-1 \, k},
\eeq
where
\bea
S_{i \, k}={\theta(n+1-k) \over x_{k-1}} \left[ \delta_{i \, 1} N_{1}^{N}
+ \theta(i-2) \theta(k-i) N_{i}^{N} \right]  - \sum_{j=i-1}^{n}{
1 \over x_{j}^{2}}\left[ y \delta_{i 1}
\delta_{k \, n+2} N_{1}^{N}+x_{i-2} \delta_{k \, i-1}
\theta(i-2) N_{i}^{N} \right] \nn
\eea
with
\beq
N^{N}_{1}={1 \over \sqrt{{\displaystyle \sum_{k=0}^{n} {1 \over
x_{k}^{2}}}\left(1+
y^{2}{\displaystyle \sum_{j=0}^{n} {1 \over x_{j}^{2}}} \right)}}, \qquad
N^{N}_{i}=
N^{C}_{i} \qquad i\geq 2.
\eeq

If we introduce these results into the general expression of $\Delta
\rho_{\rm NC}$ displayed in eq.(\ref{delta}), the latter greatly
simplifies since now
$P_{j}$, $N_{1 \, j}^{\pm}$ and $N_{2 \, jk}^{\pm}$ are
constants, independent of $j,k$. It is then immediate to find
the extra conditions necessary to get a vanishing $\Delta
\rho_{\rm NC}$.

a) The projection of the bidoublet
fields on the physical set of Goldstones and Higgses should satisfy, for fixed
$l$ and any value of $s_1$:
\beq \label{exc}
{\cal A}_{s_1 \, l+1}^{2}={\cal A}_{s_2 \, l+1}^{2},
\eeq
where $l=n+1,...,n(n+2)-1$, with the constraint that the 
sign of ${\cal A}_{s_1 \, l+1}/{\cal A}_{s_2 \, l+1}$ should be
the same for the whole column $l$. Notice that the constraint 
of eq.(\ref{exc}) is also
satisfied in this configuration when $l=0,...,n$. A final important remark
concerning this new constraint is that it
corresponds to $n(n+1)(n(n+1)-1)/2$ independent equations, exactly the
same
number of free scalar mixing angles as we have. Again, as  happens
for $n=1$,  this condition  fixes the value of these angles.

b) A second condition concerns, as in case I, the required degeneracy of
the masses of the scalar particles in order to cancel $\Delta \rho_{\rm
NC}$:
\bea \label{maseq1}
m_{l}^{\cal A}&=&m_{l}^{\cal B}
~~~~~~~~~~{\rm if}~~~~{\cal A}_{s_2~l+1}/{\cal A}_{s1~l+1}=+1
\eea
and
\bea \label{maseq2}
m_{l}^{\cal A}&=&m_{l}^{\cal C}
~~~~~~~~~~{\rm if}~~~~{\cal A}_{s_2~l+1}/{\cal A}_{s1~l+1}=-1
\eea
where $l=n+1,...,n(n+2)$.
Notice that no restriction is imposed on the masses $m_{j}^{\cal B}$
with $j=0,...,n$ ($m_{0}^{\cal B}$ is the SM-Higgs-like).

The conditions of case II are more restrictive than those of case I.
In order to obtain the screening, here, we need not only to 
put constraints on  the VEVs eq.(\ref{vevcon}) but also to fix the
common mixing angles in
the
matrices ${\cal A}$, ${\cal B}$ and ${\cal C}$ eq.(\ref{exc}), without any
arbitrariness left. Also, as in the previous case, a constraint on
the masses eq.(\ref{maseq1}) and (\ref{maseq2}) is required.

Finally the spectrum of gauge boson masses turns out to be very simple in
this
second configuration,

\bea \label{caseiimas}
M_{W_{0}^P}^{2}&=&{g_{0}^{2} \over 4} v_{0}^{2}  \qquad
M_{Z_{0}^P}^{2}={g_{0}^{2} \over 4} v_{0}^{2} \left( y^{2} \sum_{s=0}^{n}
{1 \over x_{s}^{2}} + 1 \right)
\nn
\\
M_{W_{i}^P}^{2}&=&M_{Z_{i}^P}^{2}={g_{0}^2 \over 4} \left( v_{0}^{2}+
{\displaystyle \sum_{l=0}^{i-1}} x_{l+1}^{2} u_{l \, l+1}^{2}+
x_{i-1}^{2} x_{i}^{2} u_{i-1 \, i}^{2} \sum_{r=i}^{n} {1 \over x_{r}^{2}}
\right) \qquad {i>0}.
\eea
\bigskip
It implies that,  at tree level, $\rho$ is not 1 but
\beq \label{delt}
  \rho^{tree}=1 - t_{W}^{2}\sum_{s=1}^{n}{1 \over
x_{s}^{2}},
\eeq
where $t_{W}$ stands for the tangent of the weak mixing angle.
As a  consequence of eq.(\ref{delt}) it is clear that the model requires
the extra gauge couplings to be large (as in \cite{FP})
so as to obtain an acceptable value of $\rho$ at tree level.
However, large couplings do not necessarily imply large effects in the
observables. It was noted in \cite{FP} that all the potentially large
couplings in the charged sector are always suppressed by a small mixing
angle (see eq.(\ref{crot}) in the limit of large coupling with
$x_{i}>x_{j}$
when $i>j$). On the other hand, in the neutral sector 
the terms proportional to a large coupling constant cancel in 
the sum of
diagrams with light gauge bosons and heavy scalars together with the
diagrams with only heavy gauge fields, 
leaving only subleading corrections to $\Delta \rho_{\rm NC}$.    
Similar scenarios were described in \cite{CNL} and \cite{Schilk}.

\bigskip
\subsubsection { \bf  The $\epsilon$ parameters: tree level}
\bigskip
 
Given the interesting spectrum of masses and properties of this second
case, it is worthwhile
to analyze it in more detail and compute the full set of $\epsilon$
parameters\cite{barb}. We will follow 
closely the discussion in \cite{CL} and the conventions for the parameters
$\Delta {\tilde \rho}$, $\Delta r_W$ and $\Delta k$. 
Similarly to what happen in
\cite{CL} also our model exhibits a decoupling
property in all $\epsilon$ parameters.

At tree level the Fermi constant $G_{F}$ is given in our model in case II
by
\bea  \label{Gf}
{G_{F} \over \sqrt{2}}={\pi \alpha \over 2 s_{W}^{2}}{  1 \over
\displaystyle \sum_{s=0}^{n}{1 \over x_{s}^{2}}} \left( {1 \over M_{W_{0}^P}^{2}} +  
 {\displaystyle \sum_{s=1}^{n}{1 \over x_{s}^{2}} \over M_{W_{1}^P}^{2}} \right)=
{1 \over 2 v_{0}^{2}} \left(\displaystyle 1 - \sum_{s=1}^{n} {1\over x_{s}^{2}}
\left(1 - {M_{W_{0}^P}^2 \over M_{W_{1}^P}^{2}} \right) \right)
+ {\cal O}\left({1 \over x_{i}^{4}}\right)
\eea
where  the first term in the large $x_{i}$ expansion ($1/2 v_{0}^{2}$) would 
give the exact result  if  $M_{W_{0}^P}=M_{W_{1}^P}$.

Following \cite{CL} we can also relate the cosinus of the weak mixing angle
of our model with the corresponding one in the Standard Model denoted by
$c_{0}$ in the limit of large $x_{i}$
\bea \label{com}
c_{W}^{2}=c_{0}^{2} \left(1 - t_{W}^{2} \displaystyle \sum_{s=1}^{n}{1 \over
x_{s}^{2}}+ \sum_{s=1}^{n} {1\over x_{s}^{2}}
\left(1 - {M_{W_{0}^P}^2 \over M_{W_{1}^P}^{2}} \right) \right)
\eea
where
\bea \label{csm}
c_{0}^{2}={1 \over 2} + \sqrt{{1 \over 4} - {\pi \alpha \over \sqrt{2} G_{F}
M_{Z_{0}^P}^{2}}}
\eea
Comparing the SM definition of $\Delta r_W$
\bea
{M_{W_{0}^P}^{2} \over M_{Z_{0}^P}^{2}}=c_{0}^{2} \left(1 - {s_{W}^{2} 
\over c_{W}^{2} -s_{W}^{2} } \Delta r_{W}
\right)
\eea
with eq.(\ref{caseiimas}) and eqs.(\ref{Gf}-\ref{csm}),
we find
\bea \label{derw}
\Delta r_{W}={2 \over c_{W}^{2}}{(c_{W}^{2}-s_{W}^{2}) \displaystyle 
\sum_{s=1}^{n} {1\over
x_{s}^{2}}} - {(c_{W}^{2} - s_{W}^{2}) \over s_{W}^2}
\sum_{s=1}^{n} {1\over x_{s}^{2}}
\left(1 - {M_{W_{0}^P}^2 \over M_{W_{1}^P}^{2}} \right) 
\eea
where we have used that $M_{Z_{0}^P}$ of eq.(\ref{caseiimas}) can be rewritten
in a more convenient way
\bea
M_{Z_{0}^P}^{2}={g_{0}^{2} v_{0}^{2} \over 4 c_{W}^{2}} \left(1 + y^2
\displaystyle \sum_{s=1}^{n} {1 \over x_{s}^{2}} \right)
\eea
In order to obtain $\Delta k$ and $\Delta {\tilde \rho}$ (current) we will 
compare with the standard
definition of the $Z$ current 
\bea \label{jz0}
J^{\mu}_{Z_0^P}={e \over s_{0} c_{0}} \left(1 + 
{\Delta {\tilde \rho} \over 2} \right) \sum_{i} \left[
T_{3L}^{i} - s_{0}^{2} (1 + \Delta k) Q^{i} \right] {\bar f_{i}}
\gamma^{\mu} f_{i} 
\eea
where we  denote with $f_i \equiv \{ e_L,e_R,\nu_L,u_L,u_R,d_L,d_R \}$
the chiral projections of the fermion fields
and we leave implicit
the generation indices.

From the neutral-current(NC) gauge interactions of fermions 
\bea
{\cal L}_{NC} = \left( \begin{array}{ccccc} J_{W^3_0}^{\mu } & 0 & \cdots &
&  J_B^{\mu} \end{array} \right) \left( \begin{array}{c}
W_{0\mu  }^3 \\ \vdots \\
W_{n\mu }^3 \\ B_{\mu} \end{array} \right)
\nn
\eea
\bea
= \left( \begin{array}{cccc} J_{em}^{\mu} & J_{Z_0}^{\mu  } &
\cdots & J_{Z_n}^{\mu } \end{array} \right) \left( \begin{array}{c}
A_{\mu} \\ Z_{0\mu} \\ \vdots \\ Z_{n\mu} \end{array} \right)
= \left( \begin{array}{cccc} J_{em}^{\mu} & J_{Z_0^P}^{\mu} &
\cdots & J_{Z_n^P}^{\mu} \end{array} \right) \left( \begin{array}{c}
A_{\mu} \\ Z_{0\mu}^P \\ \vdots \\ Z_{n\mu}^P \end{array} \right) \, ,
\eea
one finds that the Z current in our model is given by
\bea \label{zcuro}
J^{\mu}_{Z_0^P}=(R_N^T)_{2\,2} J_{Z_0}^{\mu} + {\displaystyle \sum_{s=1}^n
} (R_N^T)_{2+s\,2} J_{Z_s}^{\mu }
\eea
that can be further decomposed using eq.(\ref{matuik}) in
\bea \label{jz0n}
J_{Z_0}^{\mu}&=&c_{W} J_{W^3_0}^{\mu }- {s_{W}^2 \over c_{W}}{1 \over y}J_B^{\mu} 
\nn
\\
J_{Z_n}^{\mu}&=&-{P(x_{n+1}) P(x_s) \over f(s) f(s+1)} J_B^{\mu}
\eea
where 
\bea \label{j0d}
J_{W^3_0}^{\mu }=g \sum_{i} T_{3L}^{i}  {\bar f_{i}}
\gamma^{\mu} f_{i}  \qquad {\rm and} \qquad J_{B}^{\mu }={\tilde g}
\sum_{i} {\bar Y}^{i}  {\bar f_{i}}
\gamma^{\mu} f_{i} 
\eea
Comparing eq.(\ref{zcuro}-\ref{j0d}) and eq.(\ref{jz0}) one finally obtains
\bea
1+{\Delta {\tilde \rho} \over 2}&=&
\left(1 - y^{2} \sum_{s=1}^{n} {1 \over x_{s}^{2}}
+
\sum_{s=1}^{n} {1\over x_{s}^{2}}
\left(1 - {M_{W_{0}^P}^2 \over M_{W_{1}^P}^{2}} \right)
 \right) \left( (R_N^T)_{2\,2} + \sum_{s=1}^{n} {c_{W} P(x_0)^2 (R_N^T)_{2+s\,2}
 \over x_{s} f(s) f(s+1)} \right)
 \nn
\\
\Delta k &=&\sum_{s=1}^{n}{c_{W}^3 \over s_{W}^2 x_s f(s) f(s+1)} {P(x_0)^2 
(R_N^T)_{2+s\,2} \over \left( (R_N^T)_{2\,2} + \displaystyle
\sum_{s=1}^{n} {c_{W} P(x_0)^2 (R_N^T)_{2+s\,2}
 \over x_{s} f(s) f(s+1)} \right) }
\eea
where $(R_N^T)_{2+s\,2}$ can be derived from eq.(\ref{matuik}) and
eq.(\ref{rnet})
\bea
(R_N^T)_{2+s\,2}=N_{1}^N {f(-1)^2 \over f(s-1) f(s) x_s}
\eea
In the large $x_i$ expansion one finally gets
\bea \label{dkap}
\Delta {\tilde \rho}&=&-{1 \over c_{W}^2} \sum_{s=1}^n {1 \over x_{s}^2}
+ 2 \sum_{s=1}^{n} {1\over x_{s}^{2}}
\left(1 - {M_{W_{0}^P}^2 \over M_{W_{1}^P}^{2}} \right)
\nn
\\
\Delta k&=&\sum_{s=1}^n {1 \over x_{s}^2}
\eea
Recalling the definition for the $\epsilon$ parameters 
in terms of the oblique corrections $\Delta {\tilde \rho}$, $\Delta k$ and
$\Delta r_{W}$ 
\cite{barb} 
\bea
\epsilon_{1}&=&\Delta {\tilde \rho} \nn \\
\epsilon_{2}&=&c_{W}^{2} \Delta {\tilde \rho} + {s_{W}^{2}  \over
c_{W}^{2} -s_{W}^{2}}\Delta r_{W} - 2 s_{W}^{2} \Delta k \nn \\
\epsilon_{3}&=&c_{W}^{2}  \Delta {\tilde \rho} + (c_{W}^{2}-s_{W}^{2}) \Delta k
\eea
and using eq.(\ref{derw}) and eq.(\ref{dkap}) we obtain
\bea \label{epsilon79}
\epsilon_{1}&=&-{1 \over c_{W}^2} \sum_{s=1}^n {1 \over x_{s}^2}
+ 2 \sum_{s=1}^{n} {1\over x_{s}^{2}}
\left(1 - {M_{W_{0}^P}^2 \over M_{W_{1}^P}^{2}} \right)
\nn \\
\epsilon_{2}&=&{1 \over 2} \sum_{s=1}^n {1 \over x_{s}^2}
{(c_{W}^2-s_{W}^2) \over c_{W}^2} \left[ -3 + (c_{W}^2-s_{W}^2) + 
2 \left(1 - {M_{W_{0}^P}^2 \over M_{W_{1}^P}^{2}} \right) c_{W}^2 \right]
\nn \\
\epsilon_{3}&=&2 \sum_{s=1}^n {1 \over x_{s}^2}
\left[-s_{W}^2+ \left(1 - {M_{W_{0}^P}^2 \over M_{W_{1}^P}^{2}} \right) c_{W}^2 
\right].
\eea
We find that similarly to what happens in the model described in \cite {CL}
(and, of course, also in \cite{FP}) 
there is a decoupling property in the whole set of $\epsilon$ parameters
in the large $x_i$ limit.

\section { \bf  Links with Bess and degenerate Bess models}

It has been shown in \cite{FP} that the  FP model ($n=1$) in the
configuration II ($v_{1}^{(01)}=v_{2}^{(01)}=v,v_{1}=v_{0}/x_{1}$)
is able to
reproduce the masses and mixings of the gauge boson particles of the Bess
model \cite{bess} in a subset of the parameter space of both theories.

The translation table in our present notation is
\bea
&\alpha &\rightarrow \quad 1/(1+2x_{1}^{2})
\nn \\
&f &\rightarrow \quad v\sqrt{1+2 x_{1}^{2}}
\nn \\
&g^{''} &\rightarrow \quad 2 g_{0} x_{1}
\nn \\
&g &\rightarrow \quad g_{0}
\nn \\
&g^{\prime} &\rightarrow \quad {g_{0}y}.
\eea
In the left column, we have the 5 parameters of the Bess model
(we use the notation of \cite{bess}),
which
reduce to 4 free parameters with
the extra constraint $\alpha=2g^{2}/(2g^{2}+g^{''\,2})$. On the right,
we
have the 4 free parameters ($g_{0},x_{1},y,v$) of our model after
imposing the extra constraint $v_{1}=\sqrt{2} v$. In this
particular configuration the model shares all the good properties of
the Bess model plus a screening of quadratic scalar contributions to
$  \rho_{\rm NC}$.

The link with the degenerate Bess \cite{CNL}, on the contrary, turns
out to be
more complicated, and we need to impose several constraints on both
models.
Beside the presence of an
extra family of gauge bosons, the degenerate Bess model has also two important
differences with respect to the Bess model:

i) while the Bess does not decouple in the large mass limit,
the degenerate Bess does; 

ii) the $\rho$ parameter at tree level (eq.(\ref{altrerho})) 
is different from 1
in the Bess model but equal 1 in the degenerate Bess.

The degenerate Bess model has, concerning the gauge sector, 5
parameters ($r$, $x$, ${\tilde g}$, ${\tilde \theta}$ or ${\tilde
g^{\prime}}$, $v$) in the notation of \cite{CNL}.
Our model has on the contrary 13 free parameters in case I (although
some
of  them become infinitely large), and 7 free
parameters in case II.

It is not difficult to see that case I cannot overlap with
the degenerate Bess model from the condition that $M_{W_{2}}=M_{W_{1}}$.
This is equivalent to imposing that 
$x_{2}^{2} v_{2}^{2}=x_{1}^{2} v_{1}^{2}$, 
which is inconsistent in the large $v_i$ limit
with the necessary hierarchy found in case I. 
Even if one imposes an unnatural inverse hierarchy for the
couplings, the rotation matrices of case I are no longer valid, breaking
the alignment condition of eq.(\ref{cond}).
On the contrary, case II fulfills this condition automatically.

The imposed degeneracy between the new charged gauge bosons is translated
into strong constraints in both models. In the degenerate Bess we need
to impose
\beq \label{deg1}
x \rightarrow 0.
\eeq
This condition implies a complete degeneracy of all new  $Z$
and $W$ gauge bosons.
On the other side, apart from the constraints of case II,
i.e.
\bea
&&v_{1}^{(ij)}=v_{2}^{(ij)} \qquad \qquad {\rm with} \, \, \, ij=01,02,12
\nn \\
&&v_{2}={v_{0} \over x_{2}} \qquad v_{1}={v_{0} \over x_{1}}  \nn \\
&&u_{02}^{2}=u_{01}^{2} {x_{1}^{2} \over x_{2}^{2}}
\eea
one should require in addition, in our model:
\bea \label{largex}
 u_{12}^{2}={u_{01}^{2} \over x_{2}^{2}} \qquad
 x_{2} \gg x_{1} \gg 1,
\eea
that is the corresponding of condition (\ref{deg1}). The
second condition of eq.(\ref{largex})
is necessary  to ensure the absence of mixing between $W_{2}^{\pm}$
and the others $W$ gauge bosons and also to get a tree-level $\rho$
parameter equal to 1.

The translation table of the remaining parameters of the degenerate Bess
in the small-$r$ limit is
\bea \label{trans2}
&& {\tilde g} v \rightarrow g_{0} v_{0}
\nn \\
&& {\tilde c_{\theta}} \rightarrow {1 \over \sqrt{1+y^2}}={\tilde c_{W}}
\nn \\
&& r \rightarrow {v_{0}^{2} \over v_{0}^{2} + u_{01}^{2} (1 + x_{1}^{2})
},
\eea
where all terms of order $1/x_{i}^{2}$ ($i\ge 1$) have been thrown away,
and ${\tilde c_{W}}$ stands for the large $x_{i}$ ($i=1,2$) limit of
eq.(\ref{weinb}) according to eq.(\ref{largex}).
The gauge boson masses are, according to eqs.(\ref{caseiimas}),
(\ref{largex}) and (\ref{trans2}):
\bea
M_{W_{0}}^{2}&=& {g_{0}^{2} \over 4} v_{0}^{2}  \qquad
M_{Z_{0}}^{2}= {g_{0}^{2} \over 4 {\tilde c_{W}}^{2}} v_{0}^{2}  \nn \\
M_{W_{1}}^{2}&=&M_{W_{2}}^{2}=M_{Z_{1}}^{2}=M_{Z_{2}}^{2}= {g_{0}^{2}
\over 4} ({v_{0}^{2} + u_{01}^{2} (1 + x_{1}^{2})}).
\eea

The new gauge boson masses are all degenerate and larger than
the $M_{W_0}$ mass in the small $r$ limit.

\section { \bf Discussion and Conclusions}

In this paper we have presented an extension of the SM based on a gauge group
$SU(2)_L\times SU(2)^n \times U(1)_Y~~~(n\geq 1)$, with a non-trivial
scalar
sector.
We have investigated in detail the $\rho_{\rm NC}$ parameter, with
particular 
care for the scalar sector.
Imposing the exact cancellation of the quadratic terms in the masses of the
scalar fields leads to what we call a ``screening'' configuration.
We have found two different possibilities for the screening to occur and we
have calculated the respective mass spectra of the new gauge bosons.
While in one case the screening phenomenon is obtained at the price of sending
all the masses of the new gauge bosons to infinity, the second one is much
more phenomenologically interesting, because a finite, relatively light
mass
spectrum is allowed.
$\rho_{\rm NC}$ severely constrains the parameter space of the model and
we
consider as a very good indication the fact that the screening occurs for
arbitrary values of $n$.
Indeed by analyzing the set of $\epsilon$ parameters we have found that 
the model in case II exhibits a decoupling property in the large $x_i$ limit.

Of course, one should take into account other constraints coming from the
precision measuraments at LEP and Fermilab.
Indeed in a previous paper \cite{FP}, 
the phenomenology of this type of fermiophobic models was analyzed in detail
in the case $n=1$.
A tree-level fit was done to the electroweak data, also the one-loop
contributions
to $R_{b}$, $\Delta \rho$ and to all the flavour-changing processes that 
could be affected
significantly by this model were computed.
In that respect this kind of fermiophobic models
(for any $n$) has a clear advantatge with respect to other gauge extensions 
due to the absence at tree level of FCNC and to the suppressed
one-loop contributions to the process
$b \rightarrow s \gamma$ and to the $B_{0}-{\bar B}_{0}$ and
$K_{0}-{\bar K_{0}}$ mixing.
All these analysis show that relatively light $W^\prime$
and $Z^\prime$ are not excluded provided that $x$ is sufficiently large.
The Tevatron data could have changed this conclusion.
However,
analyzing the production cross section of the $W^\prime$ and $Z^\prime$,
it was shown that the previous statement remains valid.
Thanks to the fermiophobic nature of the new gauge bosons,
the $W^\prime$ couples to the fermions only through the charged mixing angle
while the $Z^\prime$ has a direct coupling that scales as $1/x$.
Therefore the weakness of these couplings that induce small cross sections
allow the existence
of relatively light new gauge bosons with no contradiction with the
Tevatron data.

All these conclusions can be applied also to our model, that reduces to the
one studied in \cite{FP} in the case $n=1$.
One can see looking to eq.(\ref{epsilon79}) that the contributions
to  the electroweak observables enter in an additive way.
These contributions are organized in a tower structure, i.e. the
lowest order $W$ and $Z$ families fix the bounds on the next ones.
The inclusion of extra $W$'s and $Z$'s receives then just more
stringent bounds on the extra
coupling constants $x_{i}$ (with $i>1$), mixing angles and masses.
The good behaviour of the case $n=1$ is not affected by the introduction of
new gauge bosons; in the allowed region of the parameter space we can
reach two different limit configurations:
we can choose to have a very light new extra boson, letting all the others
to be heavy, or to have a more balanced situation, with all the bosons
in an intermediate mass range.

The keypoint is indeed, as it was 
observed in \cite{FP}, the fermiophobic nature of these models that
automatically allows to introduce a milder modification of the whole
SM phenomenology, in contrast to other kind of gauge extensions of the SM. 
More precise and quantitative predictions can be obtained, as in the case
$n=1$ \cite{FP}, only through a complete phenomenological analysis, which
is nevertheless beyond the scope of this paper.

\bigskip
\bigskip
\noindent { \Large \bf Acknowledgements}
\medskip

The authors thank Stefano Bertolini, Roberto
Casalbuoni, Emilian Dudas, Ferruccio Feruglio and Fred Jegerlehner for useful comments
and discussions. J.M. acknowledges financial support from  a Marie Curie
EC Grant (TMR-ERBFMBICT 972147).

\pagebreak
\end{document}